\documentstyle[11pt,aaspp4]{article}
\def\ie{{\it i.e.,\ }}
\def\eg{{\it e.g.,\ }}
\def\qv{{\it q.v.,\ }}

\def\etal{{\it et al.\ }}
\def\gtrapprox{\;\lower 0.5ex\hbox{$\buildrel >
    \over \sim\ $}}             
\def\lessapprox{\;\lower 0.5ex\hbox{$\buildrel < \over \sim\ $}}
\def\msol{\ifmmode {\>M_\odot}\else {$M_\odot$}\fi}
\def\pyr{\ifmmode {\>{\rm\ yr}^{-1}}\else {yr$^{-1}$}\fi}
\def\psec{\ifmmode {\>{\rm\ s}^{-1}}\else {s$^{-1}$}\fi}
\def\kms{\ifmmode {\>{\rm km\ s}^{-1}}\else {km s$^{-1}$}\fi}
\def\psqcm{\ifmmode {\>{\rm cm}^{-2}}\else {cm$^{-2}$}\fi}
\def\pcubcm{\ifmmode {\>{\rm cm}^{-3}}\else {cm$^{-3}$}\fi}
\def\hubunits{\ifmmode {\>{\rm km\ s^{-1}\ Mpc^{-1}}}\else {km
s$^{-1}$ Mpc$^{-1}$}\fi}
\def\phoflux{\ifmmode{{\rm phot\ cm}^{-2}{\rm\ s}^{-1}}\else {phot
cm$^{-2}$ s$^{-1}$}\fi}

\newcommand{\qrms}{ Q_{\scriptscriptstyle {\rm RMS}} }
\newcommand{\tlg}{ t_{\scriptscriptstyle {\rm LG}} }
\newcommand{\sz}{ {{\rm sz}} }
\newcommand{\rmw}{ r_{\scriptscriptstyle {\rm MW}} }
\newcommand{\COBE}{{\sl COBE}}
\newcommand{\kev}{ {\rm keV} }

\def\Tkev{\ifmmode{T_{\rm kev}}\else {$T_{\rm keV}$}\fi}
\def\Em{\ifmmode{{\cal E}_m}\else{${\cal E}_m$}\fi}
\def\Dm{\ifmmode{{\cal D}_m}\else{${\cal D}_m$}\fi}

\def\be{\begin{equation}}
\def\ee{\end{equation}}
\def\bea{\begin{eqnarray}}
\def\eea{\end{eqnarray}}

\lefthead{Maloney\& Bland-Hawthorn}

\begin{document}

\title{Warm Gas and Ionizing Photons in the Local Group}

\author{Philip R. Maloney\altaffilmark{1} and
J. Bland-Hawthorn\altaffilmark{2}} 

\altaffiltext{1}{Center for Astrophysics and Space Astronomy,
University of Colorado, Boulder, CO 80309-0389; maloney@casa.colorado.edu}
\altaffiltext{2}{Anglo-Australian Observatory, P.O. Box 296,
Epping, NSW 2121, Australia; jbh@aaoepp2.aao.gov.au} 

\begin{abstract}
Several lines of argument suggest that a large fraction of the baryons
in the universe may be in the form of warm ($T\sim 10^5-10^7$ K)
gas. In particular, loose groups of galaxies may contain substantial
reservoirs of such gas. Observations of the cosmic microwave
background by \COBE\ place only weak constraints on such an intragroup
medium within the Local Group. The idea of a Local Group corona dates
back at least forty years (Kahn \& Woltjer 1959). Here we show that gas
at $T\sim 2-3\times 10^6$ K (the approximate virial temperature of the
Local Group) -- extremely difficult to observe directly -- can in
principle radiate a large enough flux of ionizing photons to produce
detectable H$\alpha$ emission from embedded neutral clouds.  However,
additional constraints on the corona -- the most stringent being
pulsar dispersion measures towards the Magellanic Clouds, and the
timing mass -- rule out an intragroup medium whose ionizing flux
dominates over the cosmic background or the major Local Group
galaxies. A cosmologically significant coronal gas mass could remain
invisible to H$\alpha$ observations. More massive galaxy groups could
contain extensive coronae which are important for the baryon mass and
produce a strong, local ionizing flux.
\end{abstract}


\section{Introduction} 
The standard Big Bang cosmological model makes remarkably precise
predictions for the abundance of baryons in the universe: in terms of
the critical density parameter $\Omega$, the prediction is
$\Omega_b\simeq(0.068\pm 0.012)h^{-2}$, where we take the present-day
Hubble constant to be $H_0=100h\hubunits$ (\eg Olive \etal 1991;
Schramm \& Turner 1998). An inventory of baryons observed
at high redshift ($z\sim 2-3$), chiefly in the form of the low-column
density Ly$\alpha$-forest clouds, gives an estimate of $\Omega_b$
which, although subject to substantial, systematic uncertainties,
is in reasonable agreement with the standard prediction of Big Bang
nucleosynthesis (see the summary in Fukugita, Hogan, \& Peebles
1998). However, as has been noted repeatedly (\eg Persic \& Salucci
1992; Fukugita \etal 1998), at $z\approx 0$ only a small fraction of
the expected number of baryons has been observed, suggesting that
there is a substantial, even dominant reservoir of baryons which has
not yet been characterized.

A plausible suggestion for one reservoir of baryons is that loose
groups of galaxies contain substantial masses of warm ($T < 10^7$ K)
ionized gas, an idea which appears to have originated with Kahn \&
Woltjer (1959; see also Oort 1970; Hunt \& Sciama 1972).  X-ray
observations of poor groups of galaxies frequently detect intragroup
gas at $T\sim 1\;\kev$ (\eg Pildis, Bregman, \& Evrard 1995; Mulchaey
\etal 1996). In general, only groups dominated by ellipticals are
detected; spiral-rich groups tend to show only emission from
individual galaxies. Although this may be
due to the absence of gas in such groups, it is also plausible that
the gas has not been seen because its temperature is too low:
the velocity dispersions characterizing groups dominated by spiral
galaxies are significantly smaller than those of compact,
elliptical-dominated groups, and imply temperatures $T\sim 0.2-0.3$
keV (Mulchaey \etal 1996), making detection even at relatively soft
X-ray wavelengths very difficult.

Most recently, Blitz \etal (1998) have suggested that the majority of
high-velocity clouds (HVCs; for a review, see Wakker \& van Woerden
1997) are not associated with the Galactic ISM, but represent remnants
of the formation of the Local Group (LG), as material continues to
fall into the LG potential. In this scenario, some fraction of these
infalling clouds will collide in the vicinity of the LG barycenter and
shock up to the virial temperature, $T\sim 2\times 10^6$ K, producing
a warm intragroup medium.

In this paper, we explore the possibility that the Local Group
contains such a reservoir of warm ionized gas. In particular, we
examine whether significant constraints can be placed on the amount of
gas through the detection of recombination lines from neutral gas
within the Local Group. In the next section we briefly recapitulate
the existing constraints on such an intragroup medium, and in \S 3 we
estimate the flux of ionizing photons. \S 4 discusses the implications
and additional constraints which can be imposed, in particular, mass
flux due to cooling and the timing mass of the Local Group.

\section{{\sl COBE} and X-Ray Constraints on Local Group Gas}

Suto \etal (1996) suggested that a gaseous LG halo could
significantly influence the CMB quadrupole moment observed by \COBE.
Assume the Local Group contains an isothermal plasma at
temperature $T_e$ whose electron number density is (for core density
$n_o$ and core radius $r_o$)
\begin{equation}
\label{eq:ner}
n_e(r) = n_o {r_o^2 \over r^2 + r_o^2}\pcubcm;
\end{equation}
\ie the nonsingular isothermal sphere.
Since we allow $r_o$ as well as $n_o$ to vary, the parameterization of
equation (\ref{eq:ner}) includes density distributions ranging from
$n_e\approx$ constant to $n_e\propto r^{-2}$. 
As in Suto \etal (1996), we calculate the resulting Sunyaev-Zeldovich
temperature decrement as a function of angle, expand in spherical
harmonics and average over the sky to obtain the monopole and
quadrupole anisotropies.
The {\sl COBE} FIRAS data (Fixsen \etal 1996) imply that the Compton
$y$-parameter $|y|= T_{0,\sz}/2 < 1.5\times10^{-5}$ (95\% CL), which
imposes the constraint
\begin{equation}
\label{eq:t0limit}
n_o r_o T_{\rm keV} < 7.4\times10^{21}\theta_o^{-1}{R\over r_o}
 \left({y \over 1.5\times10^{-5}}\right)\;\psqcm,
\end{equation}
where $\theta_o \equiv \tan^{-1}(R/r_o)$. Similarly,
the \COBE\ quadrupole moment requires
\be
\label{eq:t2limit}
n_o r_o T_{\rm keV} < 1.6\times 10^{20} Q_{{\mu\rm K}}{R\over r_o}
\left[\theta_o-3\left({r_o\over R}\right)+3\theta_o\left({r_o\over
R}\right)^2\right]^{-1}\;\psqcm
\ee
where the rms quadrupole amplitude $\qrms=10^{-6}Q_{{\mu\rm K}}$ K;
the observed value $Q_{{\mu\rm K}}\approx 6$ (\eg Bennett \etal
1996). Suto \etal (1996) argued that a LG corona which satisfied
equation (\ref{eq:t0limit}) could significantly affect the quadrupole
term, as equation (\ref{eq:t2limit}) is more restrictive than
(\ref{eq:t0limit}). However, Banday \& Gorski (1996) showed
there is no evidence for a LG corona in the \COBE\ DMR skymaps.  
In addition, Pildis \& McGaugh (1996) pointed out that
the typical values of $n_o r_o T_{\rm keV}$ observed in poor groups of
galaxies, resembling the Local Group, are well below the limit
(\ref{eq:t0limit}), generally no more than a few$\;\times
10^{20}\psqcm$. Furthermore, spiral-rich groups usually reveal no
evidence for intragroup gas at X-ray energies; Pildis \& McGaugh
give upper limits of a few$\;\times 10^{19}\psqcm$ for an assumed
temperature $\Tkev\sim 1$.

Thus, although the \COBE\ constraints on a LG corona are in fact quite
weak\footnote{There is some confusion in the literature regarding the
interpretation of the COBE limits. In evaluating eq.~(9) of Suto \etal
(1996), there is no numerical fudge factor suggested by Pildis \&
McGaugh (1996). Moreover, in Fig. 4 of Suto \etal, the 6$\mu$K curve
is displaced downwards by a factor of 3.}, analogy with similar poor
groups suggests that the LG is unlikely to have a significant gaseous
X-ray corona. However, as noted in \S 1, the lower temperature
expected for the gas in spiral-rich groups significantly relaxes the
X-ray constraints on warm gas in groups similar to the LG. {\it If}
the product $n_o r_o T_{\rm keV}$ in a LG corona is typical of that
seen in more compact groups, merely at lower temperature, the mass in
baryons can still be very substantial: for the density distribution
(\ref{eq:ner}), scaling $n_o r_o T_{\rm keV}$ to $10^{20}\psqcm$, the
mass inside radius $r$ is approximately (assuming $r/r_o \gtrapprox$ a
few)
\be
\label{eq:mofrs}
M(r)\approx 7\times10^{11}\left({r_o\over 100\;{\rm kpc}}\right)^2
\left({r\over r_o}\right)\left({n_o r_o T_{\rm keV}\over
10^{20}\psqcm}\right)\left({T_{\rm keV}\over 0.2}\right)^{-1}\;\msol\;;
\ee
this could be a substantial fraction of the mass of the Local Group
(see \S 4).

Direct detection of emission from gas at such temperatures is
exceedingly difficult. Using deep ROSAT observations, Wang \& McCray (1993)
(WM) find evidence for a diffuse thermal component with T$_{\rm keV}$
$\sim$ 0.2 and $n_e \sim 1\times 10^{-2}\ x_{\rm kpc}^{-0.5}$
cm$^{-3}$ (assuming primordial gas) where $x_{\rm kpc}$ is the
line-of-sight depth within the emitting gas in kiloparsecs. In the
next section we consider an indirect method of detection: the
recombination radiation from neutral gas embedded in the corona, due
to the ionizing photon flux generated by the corona gas.

\section{Ionizing Photon Flux from a Local Group Corona}
We assume the density distribution (\ref{eq:ner}). Approximating the
surface of a cloud as a plane-parallel slab, the normally incident
flux on the inner (facing $r=0$) cloud face is
\begin{equation}
\label{eq:phi}
\phi_i(r) \approx {\pi n^2_o r_o \over(1+r^2/r_o^2)^{1.5}}\xi_i
\left[0.8+1.3(r/r_o)^{1.35}\right]\ \phoflux
\end{equation}
where $\xi_i$ is the frequency-integrated ionizing photon emissivity
and the term in brackets is accurate to 10\% for $10^{-3} \le r/r_o
\le 12$. (For $r/r_o\gtrapprox 2$, the flux on the outer face of the
cloud is insignificant.)  To calculate $\xi_i$, we have used the
photoionization/shock code MAPPINGS (kindly provided by Ralph
Sutherland). Models have been calculated for metal abundances
$Z=0.01$, $0.1$, and $0.3$ times solar, and for equilibrium and
nonequilibrium ionization. For $10^4 < T < 10^7$ K,
$3\times 10^{-15} \le \xi_i \le 3\times 10^{-14}\;{\rm phot}\,\pcubcm\psec$
sr$^{-1}$. Scaling to physical values,
\be
\label{eq:phi2}
\phi_i(r) \approx 10^4 n^2_{-3} r_{100} \left({\xi_i\over 10^{-14}}\right)
{\left[0.8+1.3(r/r_o)^{1.35}\right]\over(1+r^2/r_o^2)^{1.5}}\ \phoflux
\end{equation}
where the central density $n_o=10^{-3}n_{-3}\pcubcm$ and the core
radius $r_o=100 r_{100}$ kpc. Poor groups show a very broad range of
core radii, from tens to hundreds of kpc (Mulchaey \etal 1996), and
typical central densities $n_o\sim\;$a few$\,\times 10^{-3}\pcubcm$
(Pildis \& McGaugh 1996). 

In Fig.~\ref{phi_r0}, we plot $\phi_i$ as a function of core radius
$r_o$ for densities $n_o=(1,3,10)\times 10^{-3}$ \pcubcm, for a
metallicity $Z=0.1$ times solar; results differ by $\lessapprox 20\%$
for the other values of $Z$. The value of $\phi_i$ is evaluated at
$r=350$ kpc, the assumed distance $\rmw$ of the Galaxy from the center
of the LG (solid lines), and at $r=0$ (dashed lines). The
fluxes can be very large, exceeding $10^6\;\phoflux$. However, for
$r_o\ll\rmw$, equation (\ref{eq:phi2}) shows that the incident flux at
$\rmw$ is greatly diminished 
compared to the peak value of $\phi_i$.

At distances $r\sim 2r_o$ or less, the ionizing flux produced
by a LG corona could be large enough for detection in
H$\alpha$: the emission measure is related to the normally
incident photon flux by $\Em=1.25\times10^{-2}(\phi_i/10^4)$ cm$^{-6}$
pc. However, to produce a significant flux, $n_o$ must be so large
that the cooling time $t_c$ within $r\sim r_o$ is short, $t_c
\lessapprox 10^9$ years. Even though the LG may be, dynamically,
considerably younger than a Hubble time, such a short cooling
timescale makes it necessary to consider explicitly the fate of
cooling gas.

To estimate the mass cooling flux $\dot M$, we assume that the flow is
steady, spherical, and subsonic, and that any gradients in the
potential are small compared to the square of the sound speed. In this
case the pressure is constant, and mass conservation requires that
$\dot M=4\pi\rho v r^2$; $v$ is the inflow velocity. The cooling
radius $r_c$
is set by the condition $t_c\sim\tlg$, the Local Group age. The flow
time from $r_c$ is $t_f\sim r_c/v\sim 4\pi\rho_c r_c^3/\dot M$, 
where $\rho_c$ is the gas density at $r_c$. We assume $t_f\sim t_c$,
so that the gas has time to cool before reaching $r=0$. This sets
$v\sim r_c/\tlg$ at $r_c$.  If the cooling function $\Lambda$ does not
vary rapidly with $T$, the density and temperature within $r_c$ scale
nearly as $\rho\propto r_c/r$, $T\propto r/r_c$ (Fabian \& Nulsen
1977). We have used these scalings to calculate $\dot M$ and $\phi_i$,
including the variation of $\xi_i$ and $\Lambda$ with
radius. Fig.~\ref{phi_mdot} shows several models. The ionizing flux
can be large for small $\dot M$ if $r_o$ is large and $n_o$ is low,
but in many cases $\dot M$ is prohibitively large, ruling out any such
coronae.  However, there are several important caveats. Unless the LG
is very old, it is unlikely that a steady-state flow has been
established (\eg Tabor \& Binney 1993), especially as infall of gas
into the LG is likely to be ongoing. (If a steady-state flow existed
with substantial $\dot M$, one would expect the line luminosity -- \eg
H$\alpha$ -- from the cooled gas to be high: see Donahue \& Voit
1991.) Furthermore, $\dot M$ is sensitive to the assumed density
distribution. For a given metallicity (and therefore $\Lambda(T)$) and
LG age, there is a unique value of $n_o$ at which $t_c=\tlg$ and $\dot
M\rightarrow 0$. As $n_o$ is raised above this value $\dot M$
increases rapidly, since $r_c$ increases and $\dot M\propto
r_c^2$. The value of $\phi_i$ at a given value of $\dot M$ also
depends on $Z$, since the reduced $\Lambda$ for low $Z$ means that
$n_o$ is larger for a fixed $t_c$. Given these uncertainties, it is
not clear that the estimated values of $\dot M$ should be regarded as
serious constraints.

\section{Discussion}

The results of the previous section show that a warm Local Group
corona could in principle generate a large enough ionizing photon flux
to produce detectable H$\alpha$ emission from neutral hydrogen clouds
embedded within it. This would offer an indirect probe of gas which is
extremely difficult to observe in emission. Whether the flux seen by
clouds at distances comparable to the offset of the Galaxy from the
center of the Local Group is high enough for detection depends to a
large extent on the core radius characterizing the gas distribution,
due to the dropoff in flux for $r$ substantially greater than
$r_o$. As shown in Fig.~\ref{phi_r0}, for sufficiently large values of
$r_o$ and $n_o$, $\phi_i$ can be detectably large even at a few
hundred kpc from the LG barycenter.

These large$-n_o$, large$-r_o$ models run into insurmountable
difficulties, however, when we examine the additional constraints
which can presently be imposed on a LG corona. In Fig.~\ref{n0_r0} we
show, shaded in gray, the range in ($r_o,n_o$) for which the resulting
ionizing photon flux is between $\phi_i=10^4$ and $\phi_i=10^5$
\phoflux, for radial offsets $r=0$ (lower region) and $r=\rmw=350$ kpc
(upper region). The cosmic background is probably $\phi_{\rm i,cos}
\sim 10^4$ \phoflux\ (Maloney \& Bland-Hawthorn 1999: MBH). We 
also plot the following constraints:

\noindent (1) The assumption that any LG intragroup medium is ``typical''
(Mulchaey \etal 1996; Pildis \& McGaugh 1996)
constrains the product $n_o r_o \lessapprox 1.5\times 10^{21}\psqcm$,
assuming $\Tkev\sim 0.2$. This is plotted as
the short-dashed line in Fig.~\ref{n0_r0}. Any corona which is not
unusually rich must lie to the left of this line. This restriction
alone rules out any significant contribution to $\phi_i$ at $\rmw$.

\noindent (2) Assuming that the relative velocity of approach of the
Galaxy and M31 is due to their mutual gravitational attraction, one
can estimate the mass $M_T$ of the Local Group (Kahn \& Woltjer 1959;
\qv Zaritsky 1994). This `timing mass' depends somewhat on the choice
of cosmology; we take $M_T=5\times 10^{12}\msol$ within $r=1$
Mpc of the LG center. The timing mass constraint (using equation [10])
is shown as the solid line in Fig.~\ref{n0_r0}. As plotted, it is
barely more restrictive than the \COBE\ quadrupole constraint (the
long-dashed line), and is only more stringent than restriction (1) for
large core radii. However, realistically the $M_T$ constraint is much
more severe, as the Milky Way and M31 undoubtedly dominate the mass of
the Local Group, and so the timing mass curve in Fig.~\ref{n0_r0}
should be moved downward in density by a factor of at least $\sim
5-10$.

\noindent (3) We possess some information on (more precisely, upper
limits to) the actual electron densities at $r\sim \rmw$. Constraints
on $n_e(\rmw)$ come from two sources. Observations of dispersion
measures \Dm\ toward pulsars in the LMC and the globular cluster NGC
5024 (Taylor, Manchester \& Lyne 1993) require a mean $n_{-3}\sim 1$;
this is a slightly weaker constraint than provided by $M_T$.
However, most of this column must be contributed by the Reynolds
layer, and some fraction of the \Dm\ toward the LMC pulsars presumably
arises within the LMC, so probably $\lessapprox
10\%$ can be due to a LG corona. Second, a mean density of no
more than $n_{-3}\sim 0.1$ is allowed by models of the Magellanic
Stream; otherwise, the Stream clouds would be plunging nearly radially
into the Galaxy (Moore \& Davis 1994). This limits the central density
to $n_{-3} \approx 0.1 + (\rmw/r_o)^2$. The hatched region in
Fig.~\ref{n0_r0} indicates the portion of $(r_o,n_o)$ space in which
$n_e(\rmw)\le 10^{-4}$ \pcubcm.

\noindent (4) As noted earlier (\S 2), WM found evidence for a
thermal soft X-ray component at $\Tkev\approx 0.2$. If this emission
arises in a LG corona, then the corresponding electron density as
derived from the emission measure \Em\ is $n_e \sim 3\times 10^{-4}\
x_{\rm Mpc}^{-0.5}$ cm$^{-3}$, where $x$ is the extent of emitting
region along the line of sight; the density would be $\sim 3$ times
smaller for gas of solar rather than zero metallicity. This density
constraint is comparable to the \Dm\ constraint plotted in
Fig.~\ref{n0_r0}.

Some of these constraints can be avoided if the corona gas is
clumped. The estimates of mass (equation[\ref{eq:mofrs}]) and $\phi_i$
assume a smooth density distribution. However, if the actual densities
are a factor $C$ higher than the mean (smoothed) density at a given
radius, $\phi_i$ can be kept constant while reducing both the gas mass
and \Dm\ by $1/C$. This is {\it ad hoc}, but if the
LG halo is being fueled by ongoing infall, it would not be at all
surprising for the gas distribution to be nonuniform. However, the WM
X-ray determination is unaffected by clumping, as it is derived from
\Em.

The constraints on a LG corona shown in Fig.~\ref{n0_r0} rule out a
significant contribution to the ionizing flux at $r\sim\rmw$. If the
core density $n_o$ is high, the core radius $r_o$ must be small;
conversely, for large $r_o$, $n_o$ must be low. LG coronae within the
allowed region of parameter space can produce fluxes 
$\phi_i\gg\phi_{\rm i,cos}$, but only on scales of a few tens of
kpc, at best. Thus the maximum volume in which a corona ionizing flux
exceeds $\phi_{\rm i,cos}$ is only of order $1\%$ of the LG volume,
comparable to the volume which can be ionized by galaxies
(MBH). This has important implications for the model of Blitz \etal
(1998), in which most HVCs are remnants of the formation of the Local
Group. If HVCs are at megaparsec distances, $\phi_i$ will be dominated
by the cosmic background. The resulting emission measures will be
small: barring unusually favorable geometries, the expected H$\alpha$
surface brightnesses ($\lessapprox 10$ mR) are at the limit of
detectability. Any HVCs which are truly extragalactic {\it and}
detectable in H$\alpha$ would need to lie close to the dominant spiral
galaxies (within their ``ionization cones'': Bland-Hawthorn \& Maloney
1999a,b) or the LG barycenter.

In summary, a warm LG corona which significantly dominates the UV
emission within the Local Group is ruled out, although such a corona
could contain a cosmologically significant quantity of baryons. More
massive galaxy groups could well contain coronae that are both
cosmologically important and dominate over the ionizing
background. Such coronae could have major impact on the group galaxies
through ionization and ram pressure stripping\footnote{We note that,
in principle, observations of the O$\;$VI doublet at 1032 and 1038
\AA\ are extremely sensitive to the presence of such a corona: for the
maximum allowed coronae of Fig.~\ref{n0_r0}, the expected line fluxes
could be as large as $F\sim {\rm a\ few}\times 10^{-10}$ erg cm$^{-2}$
s$^{-1}$. However, the observational difficulties (absorption and
scattering of the photons within the ISM of the Galaxy and the very
large spatial extent of the source for a LG corona) are
severe.}. Finally, we note that, four decades later, the observational
limits on a LG corona have yet to improve on the values suggested by
Kahn \& Woltjer (1959).

PRM is supported by the Astrophysical Theory Program under NASA grant
NAG5-4061.

\clearpage
\figcaption[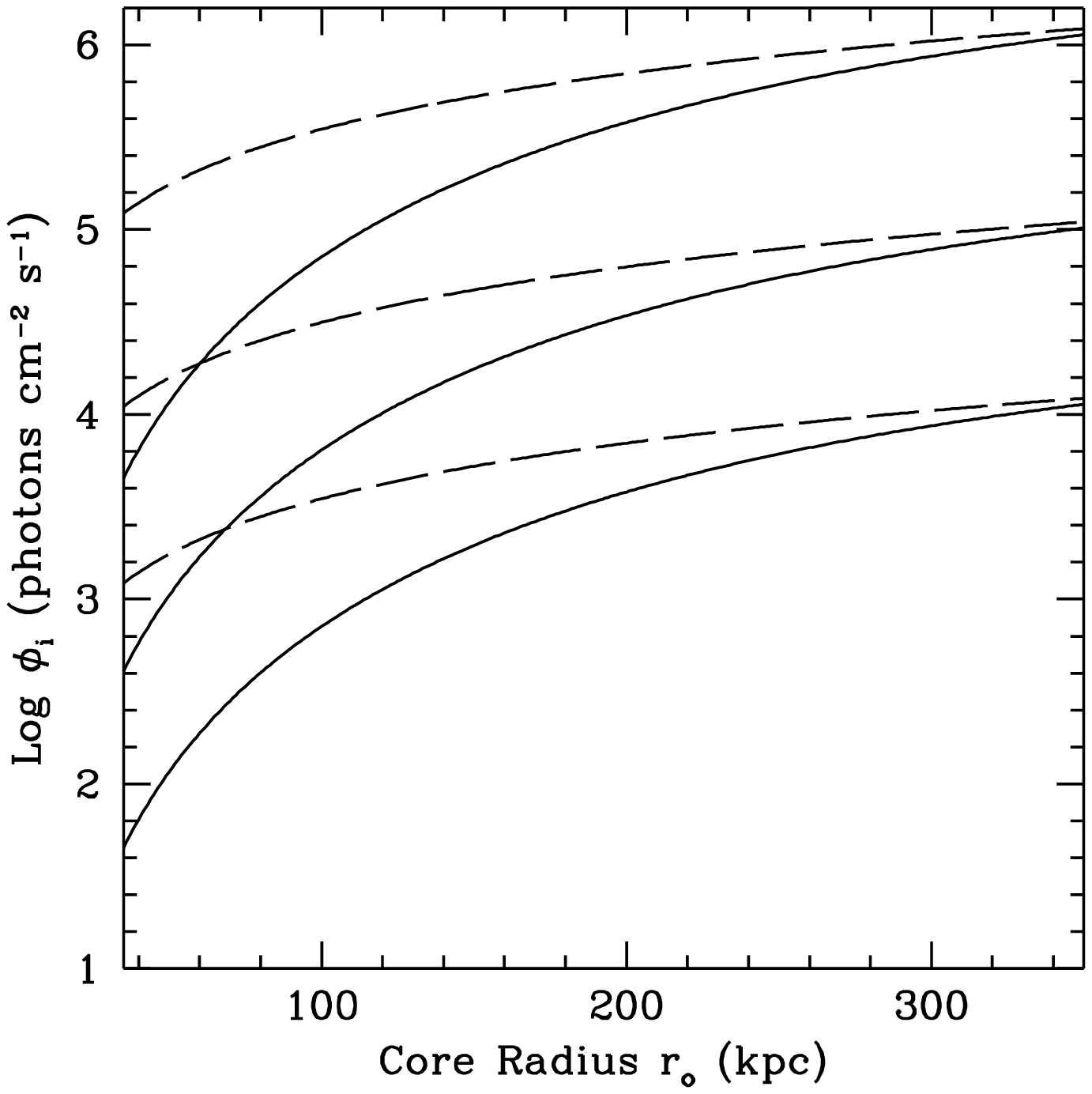]{Normally incident ionizing photon fluxes
$\phi_i$ from a Local Group corona, for core densities
$n_0=(1,3,10)\times 10^{-3}\pcubcm$ (bottom to top), as a function of
core radius $r_o$. The solid lines are for $r=350$ kpc from the center
of the corona, and the dashed lines are for $r=0$. A metallicity
$Z=0.1 Z_\odot$ has been assumed.\label{phi_r0}}


\figcaption[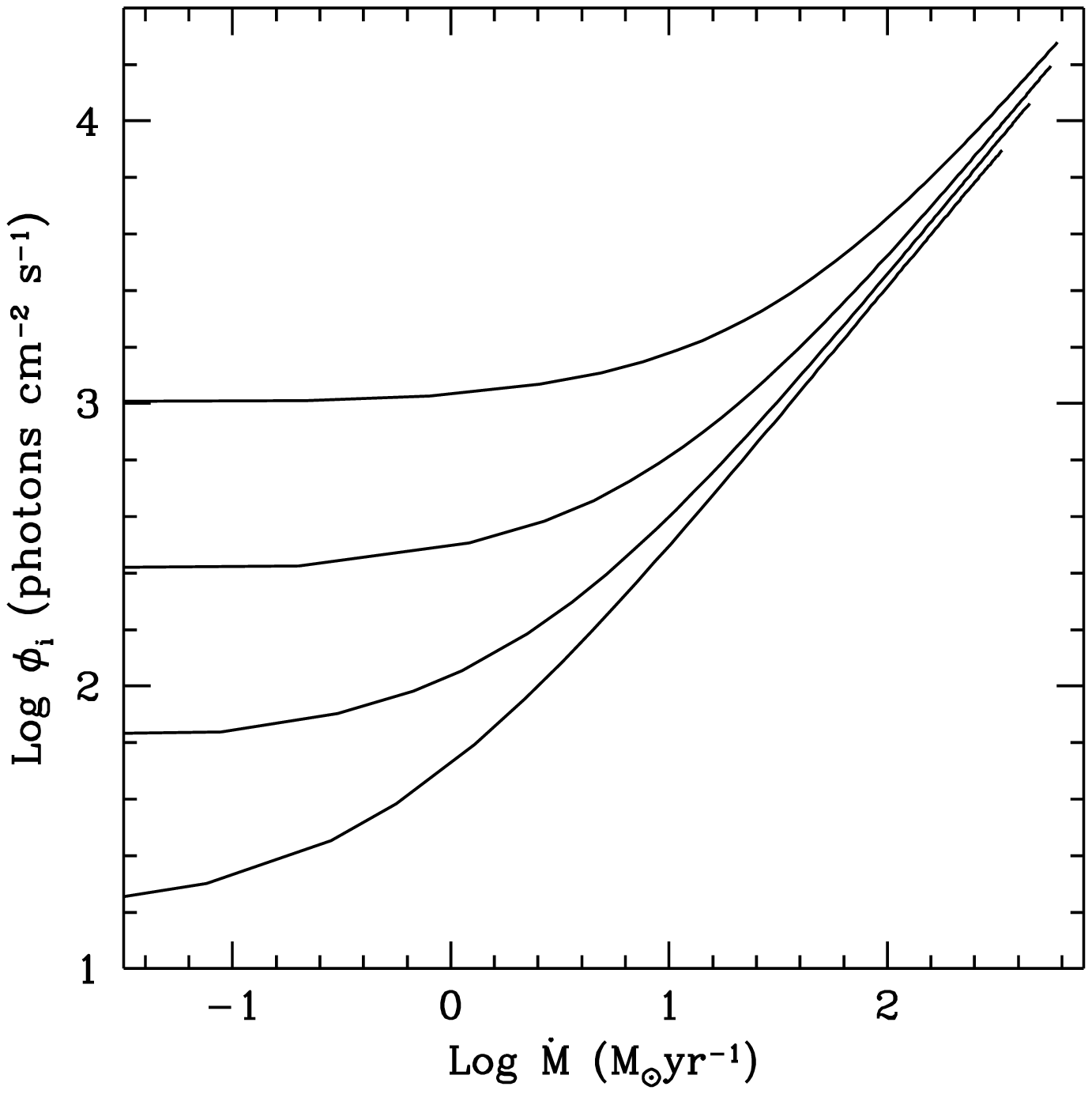]{Ionizing photon fluxes $\phi_i$ versus the
mass cooling rate $\dot M$, for a steady-state cooling flow. A core
radius $r_o=50$ kpc has been assumed. From top to bottom, the curves
are for assumed corona ages $\tlg=1$, 2, 4, and 8 Gy.\label{phi_mdot}
}

\figcaption[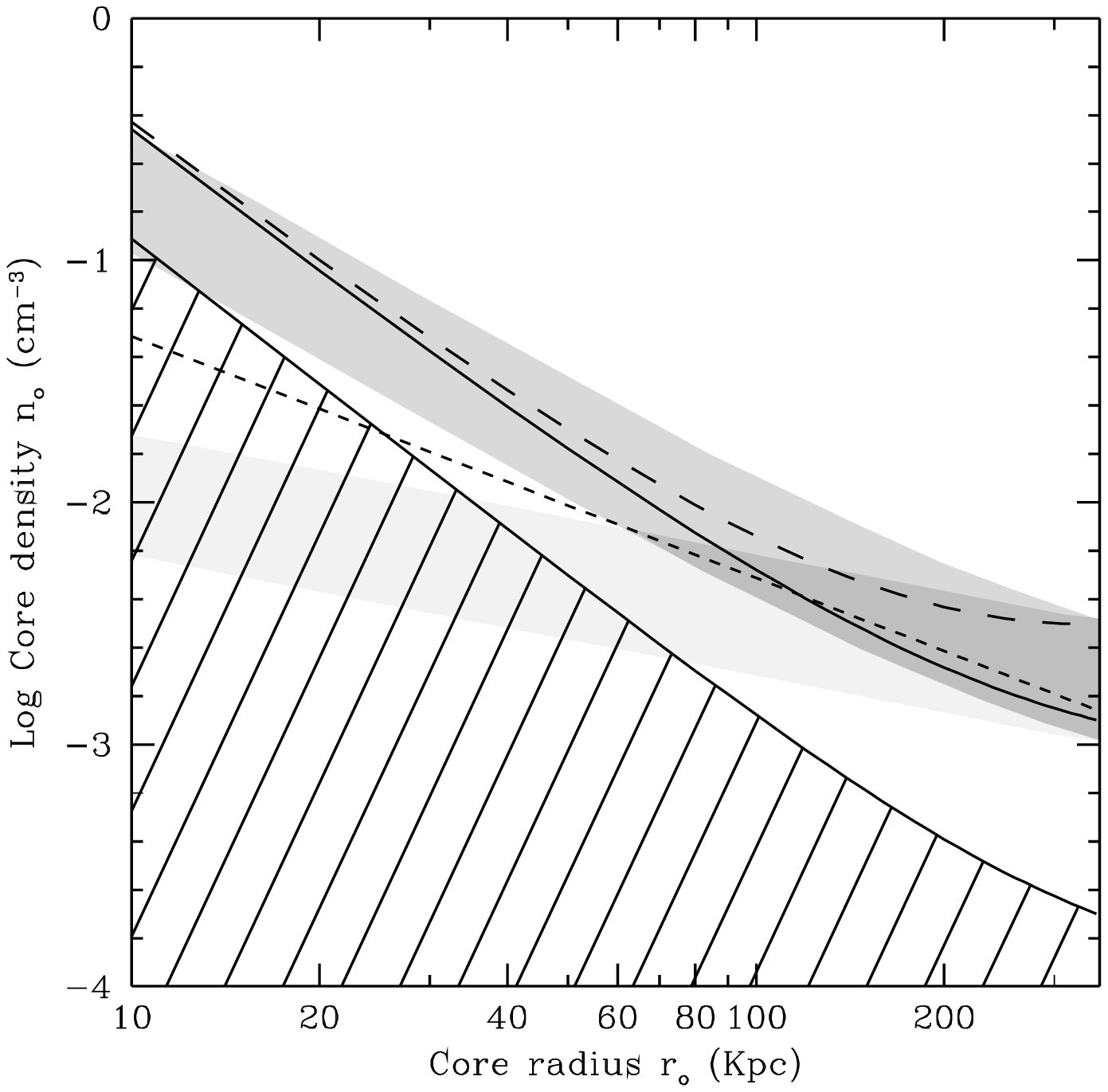]{Constraints on a Local Group corona in the
$(r_o,n_o)$ plane.  Coronae within the gray-shaded regions produce
ionizing photon fluxes between $\phi_i=10^5$ and $10^4$ \phoflux \
(upper and lower edges) at radii $r=0$ (lower region) and $r=350$ kpc
(upper region) with respect to the LG center. The long-dashed line is
the {\it COBE} quadrupole constraint, the short-dashed line assumes
the LG medium is ``typical'', the solid line is the timing mass
constraint, and the hatched region satisfies $n_e\le 10^{-4}\pcubcm$
at $r=350$ kpc. See \S 4 for discussion.\label{n0_r0}
}

\clearpage
\setcounter{figure}{0}
\begin{figure}
\plotone{fig1.eps}
\caption{}

\end{figure}


\begin{figure}
\plotone{fig3.eps}
\caption{}
\end{figure}

\begin{figure}
\plotone{fig4.eps}
\caption{}
\end{figure}

\end{document}